\documentclass[         %
aps,                    
prd,                    
preprint,               
tightenlines,          
nofootinbib,            
preprintnumbers,        %
floatfix]               
{revtex4}               
\usepackage{graphicx,longtable}

\begin{document}

\title{Unparticle Searches Through Compton Scattering}

\author{O. \c{C}ak{\i}r}\email{ocakir@science.ankara.edu.tr}
\affiliation{Department of Physics, Ankara University, 06100 Tandogan,
Ankara, Turkey}
\author{K. O. Ozansoy}\email{oozansoy@physics.wisc.edu}
\affiliation{Department of Physics, University of Wisconsin, Madison,
WI 53706, USA}
\affiliation{Department of Physics, Ankara University, 06100 Tandogan,
Ankara, Turkey}

\date{\today}

\begin{abstract}
We investigate the effects of unparticles on Compton scattering,
$e \gamma\to e \gamma$ based on a future $e^+e^-$ linear collider
such as the CLIC. For different polarization configurations, we
calculate the lower limits of the unparticle energy scale
$\Lambda_{\cal U}$ for a discovery reach at the center of mass
energies $\sqrt{s}=0.5$ TeV- $3.0$ TeV. It is shown that,
especially, for smaller values of the mass dimension $d$, ($1 <d
<1.3$), and for high energies and luminosities of the collider
these bounds are very significant. As a stringent limit, we find
$\Lambda_{\cal U}>~80$ TeV for $d<1.3$ at $\sqrt{s}=3$ TeV, and
$1$ ab$^{-1}$ integrated luminosity per year, which is comparable with the
limits calculated from the other low and high energy physics
implications.
\end{abstract}

\medskip


\maketitle

\section{Introduction}

Recently,  Georgi proposed that the possibility of new physics
effects of a hidden scale invariant sector living at a very high
energy scale\cite{Georgi:2007ek}. If such a scale invariant sector
really exists then one can expect to observe the effects of it on
our low energy physical phenomena. A very appealing feature of
Georgi's unparticle formulation for such a scale invariant sector
is that one can investigate the implications of it on the current
low and high energy physics  data. Considering the idea of
\cite{Banks:1981nn}, in \cite{Georgi:2007ek}, the scale invariant
sector is presented by a set of the Banks-Zaks operators ${\cal
O}_{BZ}$, and defined at the very high energy scale. Interactions
of ${\cal O}_{BZ}$ and the  Standard Model(SM) operators ${\cal
O}_{SM}$ are expressed by the exchange of particles with a very
high energy mass scale ${\cal M}_{\cal U}^k$ in the following form
\begin{eqnarray}
\label{1}
 \frac{1}{{\cal M}_{\cal U}^k}{O}_{BZ}{O}_{SM}
\end{eqnarray}
where BZ, and SM operators are defined as
${O}_{BZ}\in {\cal O}_{BZ}$ with mass dimension $d_{BZ}$,
and ${O}_{SM} \in {\cal O}_{SM} $ with mass dimension $d_{SM}$.
Low energy effects of the scale invariant ${\cal O}_{BZ}$ fields
imply a  dimensional transmutation. Thus, after
the dimensional transmutation Eq.(\ref{1}) is given as
\begin{eqnarray}
\label{2}
 \frac{C_{\cal U} \Lambda_{\cal U}^{d_{BZ}-d}}{{\cal M}_{\cal U}^k}{O}_{\cal U}{O}_{SM}
\end{eqnarray}
where $d$ is the mass dimension of the unparticle operator
$O_{\cal U}$ (in Ref. \cite{Georgi:2007ek}, $d=d_{\cal U}$ ), and
the constant $C_{\cal U}$ is a coefficient function.

In Refs. \cite{Georgi:2007ek} and \cite{Georgi:2007si} main ideas for
effective interactions between SM fields and the unparticles have
been presented by Georgi. In \cite{Cheung:2007ue}, the collider
phenomenology of unparticle physics has been explored in a great
detail, and Feynman rules for spin 0, spin 1, or spin 2
unparticles coupled to a variety of SM gauge invariant operators
have been explicitly given. Subsequently, many analysis on
unparticles have been done for possible signatures and for limits
from the collider experiments, astrophysical, and cosmology
implications Ref. \cite{unparticles}.

Searching for the new physics effects, the $e^+e^-$ linear
colliders have an exceptional advantageous for its appealing clean
background, and the possibility for the options of $e\gamma$, and
$\gamma\gamma$ colliders based on it.  Recently, for the new
physics searches, as a multi TeV energy electron-positron
linear collider, the CLIC is seriously taken into account, and
there are numerous works on the phenomenology potential of it. For
the multi-TeV linear electron positron colliders and the physical
potential  of CLIC one can consult \cite{collect}.
The CLIC $e^+e^-$ linear collider would also
have the options for $e^-e^-,e\gamma$, and $\gamma\gamma$ collider
options, and possibilities of polarized $e^+,e^-$ beams. The
$e\gamma$ option of a linear electron-positron collider, is also a
favorable tool for searching new physics effects for
example \cite{Dawson:2004xz,Davoudiasl:1999ni,Cakir2003}. Namely,
the center of mass energy of the $e\gamma$ collision can reach
more than $90\%$ of main $e^+e^-$ center of mass energy. And also
luminosity can be comparable with that of the main $e^+e^-$
collision. In \cite{Ginzburg:1983}, a detailed analysis on
$e\gamma$ option of an $e^+e^-$ collider has been given.

Here, we consider the CLIC based $e\gamma$ collider to search for
the unparticle physics effects, our results can easily be extended
for other possible future TeV-scale linear electron-positron
colliders. Our calculations show that the unpolarized, and
polarized $e\gamma$ collisions at TeV scale energies, and high
luminosities are very sensitive for the unparticle searches.

\section{Compton Scattering}

The Standard Model scattering amplitude  for the Compton
scattering is given as
\begin{eqnarray}
 \label{3}
M_{SM}=M^s_{SM}+M^u_{SM}
\end{eqnarray}
where
\begin{eqnarray}
\label{4}
M^s_{SM}=&&-\frac{e^2}{2p_1.k_1}
[\bar e(p_2) \displaystyle{\not} \epsilon^*(k_2) \displaystyle{\not} \epsilon(k_1)
\displaystyle{\not} k_1 e(p_1)] \\
\label{5}
M^u_{SM}=&&-\frac{e^2}{2p_1.k_2}\Big [
[\bar e(p_2) \displaystyle{\not} \epsilon(k_1) e(p_1)](p_1.\epsilon^*_2)]
+[\bar e(p_2) \displaystyle{\not} \epsilon(k_1) \displaystyle{\not} \epsilon^*(k_2)
\displaystyle{\not} k_2 e(p_1)]\Big ]
\end{eqnarray}
where $\epsilon(k_1)(\epsilon(k_2))$ is the polarization vector
of the incoming(outgoing) photon to the vertex,
$k_1(k_2)$, and $p_1(p_2)$ are the momenta of the
incoming(outgoing) photon and electron.
Therefore, using the center of mass reference frame
kinematical relations relevant for
the process $e^-\gamma\to e^-\gamma$ we get
\begin{eqnarray}
 \label{6}
|M_{SM}|^2=-2e^4[\frac{s}{u}+\frac{u}{s}]
\end{eqnarray}
where, considering the high energy limits, we take the electron mass to be zero.

\begin{figure}
\includegraphics
{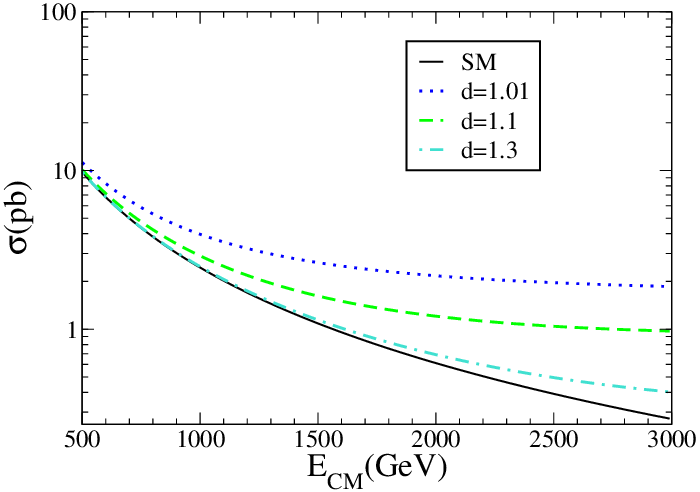} \caption{ Total unpolarized cross
sections depending on the center of mass energy of the $e^+e^-$
system for SM, and SM+${\cal U}_S$, where we take
$\Lambda_{\cal U}=2$ TeV, $\lambda_0=1$ \label{fig:xs0}}
\end{figure}

Interactions of the scalar
unparticles with the electrons, and photons are given
by \cite{Georgi:2007si,Cheung:2007ue}.
Therefore, the contribution to the scattering amplitude
from the exchange of the scalar unparticle takes the form
\begin{equation}
\label{7}
{M^t}_{{\cal U}_S}=\frac{f(d)}{\Lambda_{\cal U}^{2d-1}}
[\bar e(p_2) e(p_1)]
[(\epsilon_2^*.\epsilon_1)(k_1.k_2)-(\epsilon_2^*.k_1)(\epsilon_1.k_2)]
[-q^2-i\epsilon]^{d-2},
\end{equation}
where
\begin{equation}
\label{8-9}
f(d)=\frac{2{\lambda_{0}^2} A_d}{ \sin (d\pi)},
\quad A_d=\frac{16\pi^{5/2}}{{(2\pi)}^{2d}}
\frac{\Gamma(d+1/2)}{\Gamma(d-1)\Gamma(2d)}.
\end{equation}
In principle, the coupling $\lambda_{\gamma 0}$ of the scalar unparticle
to the photons can be different than the coupling $\lambda_{e0}$ to the electrons.
Here, for simplicity, we assume $\lambda_{\gamma 0}=\lambda_{e0}=\lambda_{0}$
without lost of generality. Thus, for the scalar unparticle exchange, one can find
\begin{eqnarray}
\label{10}
|M^t_{{U}_S}|^2=\frac{[f(d)]^2}{4\Lambda_{\cal U}^{(4d-2)}}[-t]^{2d-1}
\end{eqnarray}
In Figure \ref{fig:xs0}, we plot a schematic view of effects of
the scalar unparticles to the total unpolarized cross section with
respect to the center of mass energy $\sqrt{s}=E_{CM}$ of $e^+e^-$
collision. Plotting this figure, we assume the 
unparticle energy scale $\Lambda_{\cal U}=2$ TeV,
scalar unparticle coupling $\lambda_0=1$, and the center of mass frame scattering angle
region $|\cos\theta|<0.9$. According to the Figure \ref{fig:xs0},
one can see that, for the mass dimension values $d<1.3$
effects of the scalar unparticles are very significant.

\begin{figure}
\includegraphics
{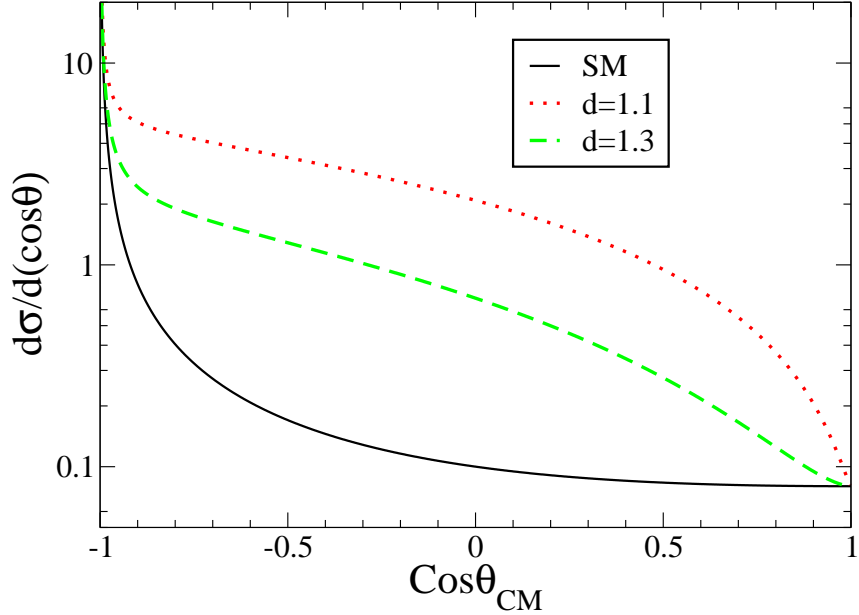} \caption{ The unpolarized
differential cross section depending on the scattering angle for
SM and SM+${\cal U}_S$ at $\sqrt{s}=3$ TeV, where we take
$\Lambda_{\cal U}=1$ TeV, $\lambda_0=1$ \label{fig:dxs0}}
\end{figure}

In Figure \ref{fig:dxs0}, we present the unpolarized differential
cross section  depending on the scattering angle for the SM and
SM+${\cal U}_S$ for $d=1.1, 1.3$. We assume $\lambda_0=1$ and
$\Lambda_{\cal U}=1$ TeV at the center of mass energy of
$\sqrt{s}=3$ TeV. One can notice that under these assumptions, the
scalar unparticle effect on the differential cross section is very
significant.

\section{Unparticle Effects on the Polarized Compton Scattering}

Polarized cross section for the Compton scattering depends on the
average helicity of the back-scattered photon
$h_\gamma=h_\gamma(h_0,h_e)$ (where $h_0$ is the helicity of the
laser photon, $h_e$ is the helicity of the initial electron), and
the helicity $P$ of the electron can be written as
\begin{eqnarray}
\label{11}
\frac{d\sigma}{d\cos\theta} = &&\frac{1}{(32\pi)}\int_{x_{min}}^{0.83}dx
\frac{f_{\gamma}(x)}{xs_{ee}}
\\
\nonumber
&& \times \Big[ \Big ( \frac{1+P h_{\gamma}(x)}{2} \Big )
\Big |M^{SM+{\cal U}_S}(RR) \Big |^2
+\Big ( \frac{1-P h_{\gamma}(x)}{2} \Big ) \Big |M^{SM+{\cal U}_S}(RL) \Big |^2\Big]
\end{eqnarray}
where $\sqrt{\hat s}=\sqrt{xs}$ is the reduced center
of mass energy of the back scattered photon beam,
x is the energy fraction, $x_{min}=E_0 E_e/m_e^2$,
$E_0$($E_e$) is the energy of laser photon(initial electron)
with respect to the laboratory frame, $m_e$ is the electron mass.
We sum over the final helicities, therefore,
$R,L$ stand for the initial helicities of the electron,
and the photon beams.
Also, the energy spectrum of the back scattered photon
$f_{\gamma}(x)$ is given as follows \cite{Ginzburg:1983}

\begin{eqnarray}
 \label{12}
f_{\gamma}(x,h_e,h_0)=\frac{1}{g(\kappa)+h_eh_0y(\kappa)}
\big [\frac{1}{1-x}+1-x-4r(1-r)+h_eh_0r\kappa(1-2r)(2-x) \big]
\end{eqnarray}

where $r=x/\kappa(1-x)$, $\kappa=4E_e E_0/m_{e}^2$, and

\begin{eqnarray}
 \label{13}
g(\kappa) &=& \big[1-\frac{4}{\kappa}-\frac{8}{\kappa^2} \big]
ln(\kappa+1)+\frac{1}{2}+\frac{8}{\kappa}-\frac{1}{2(\kappa+1)^2}
\\
y(\kappa) &=& \big[1+\frac{2}{\kappa}\big] ln(\kappa+1)
-\frac{5}{2}+\frac{1}{\kappa+1}-\frac{1}{2(\kappa+1)^2}
\end{eqnarray}

The average helicity of the back-scattered photons
is given \cite{Ginzburg:1983}

\begin{eqnarray}
 \label{14}
h_{\gamma}(x,h_e,h_0)=\frac{h_0(1-2r)(1-x+\frac{1}{1-x})+h_er\kappa[1+(1-x)(1-2r)^2]}
{1-x+\frac{1}{1-x}-4r(1-r)-h_eh_0r\kappa[(2r-1)(2-x)]}.
\end{eqnarray}

\begin{figure}
\includegraphics
{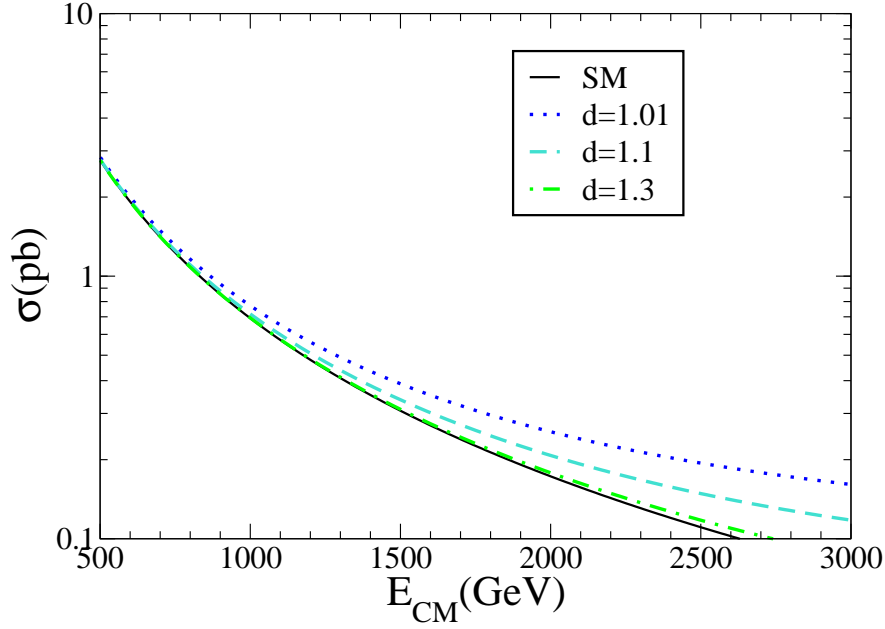} \caption{\label{fig:xsp1} Total
polarized cross sections for SM, and SM+${\cal U}_S$ with the
polarization configuration $P_1\equiv (h_e=1,h_0=1,P=-1)$.}
\end{figure}

\begin{figure}
\includegraphics
{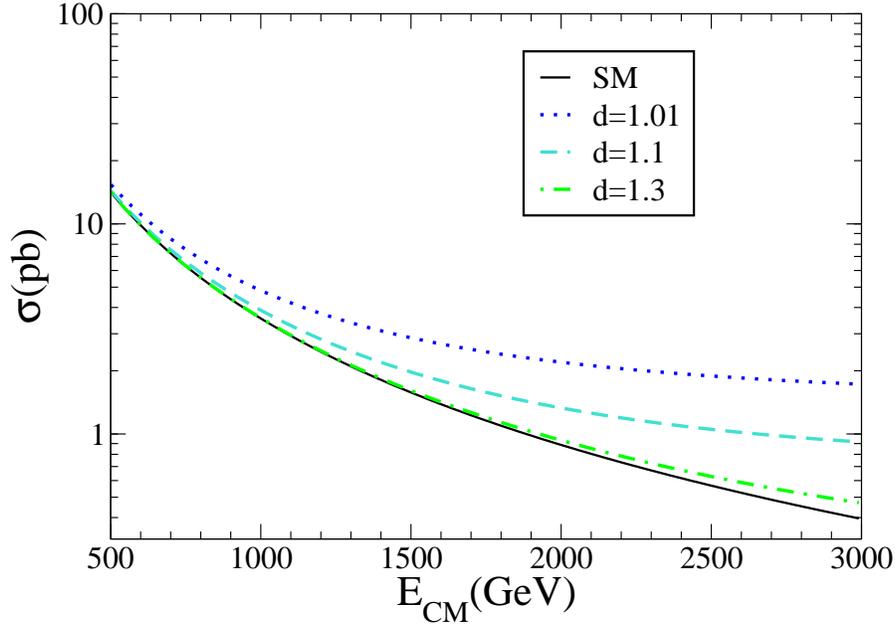} \caption{\label{fig:xsp2} Total
polarized cross sections for SM, and SM+${\cal U}_S$ with the
polarization configuration $P_2\equiv (h_e=1,h_0=-1,P=-1)$.}
\end{figure}

Using the above expressions, for different configurations of
$(h_e,h_0,P)$, we plot the total cross section with respect to the
center of mass energy $E_{CM}$. In Figure \ref{fig:xsp1}, and
Figure \ref{fig:xsp2}, we plot a schematic view of effects of the
scalar unparticles on the total polarized cross section for the
polarization configurations $P_1\equiv (h_e=1,h_0=1,P=-1)$, and
$P_2\equiv (h_e=1,h_0=-1,P=-1)$, respectively. For the figures, we
assume $\Lambda_{\cal U}=2$ TeV, $\lambda_0=1$, and the center of
mass frame scattering angle region $|\cos\theta|<0.9$. According
to the Figure \ref{fig:xsp1}, and \ref{fig:xsp2}, one can
immediately notice that for the mass dimension values
$d<1.3$, the effects of the scalar unparticles are considerably
significant. And, also one can see that, due to the polarization
dependence of the photon number density, the cross section values
associated with the polarization configurations $P_1$ and $P_2$
are remarkably different.

\section{Limits and Discussions}

In the calculations, for the discovery reach of the energy scale
of unparticles,  we use the standard chi-square analysis for the
following $\chi^2$ function

\begin{eqnarray}
\label{15}
 \chi^2=\big ( \frac{\cal L}{\sigma^{SM}}\big)
[{\sigma^{SM}-\sigma^{SM+{\cal U}}
(\Lambda_{\cal U})}]^2
\end{eqnarray}

For the one sided chi-square analysis we assume $\chi^2\geq 2.7$. And,
to show that the luminosity dependence of the unparticle energy
scale $\Lambda_{\cal U}$, we take two possible integrated luminosity values,
${\cal L}=100fb^{-1}$, and ${\cal L}=1000fb^{-1}$ per year. We
observe that due to the polarization dependence of the photon
number density, the limits are considerably different for
different polarization configurations. Here, we present the limits
for the $P_1=(h_e=1,h_0=-1,P=-1)$ polarization configuration which
gives the strongest limits comparing the other polarization
configurations. Our results is given in the TABLE \ref{tab1}.
As a remark, our lower limits on $\Lambda_{\cal U}$ are extracted
assuming  $\lambda_0=1$. In a similar calculation,
one can also find upper limits on $\lambda_0$ assuming a fixed
value of $\Lambda_{\cal U}$ such as $\Lambda_{\cal U}=1000$GeV. For example,
using our limits presented in the
TABLE \ref{tab1} for the values $d=1.5$, $\sqrt{s}=3000$GeV,
and ${\cal L}=100 (1000) \text{fb}^{-1}$ one can
find ${\lambda_0}_{max}=0.49(0.43)$
which is consistent with the limits given in the literature \cite{Anchordoqui:2007dp}.

\begin{table}
{\caption{\label{tab1}Lower limits on the $\Lambda_{\cal U}$
in the units of GeV; the first entries in the paranthesies are for ${\cal L}=100fb^{-1}$,
and the second ones for  ${\cal L}=1000fb^{-1}$}.}
\begin{ruledtabular}
\begin{tabular}{lcccccc}
$\sqrt{s}$ GeV &  d=1.0001  & d=1.01 & d=1.1 & d=1.3 & d=1.5 & d=1.9 \\
\hline
500  &$(20300, 36100)$ & $(18000, 31700)$ & $( 7400, 11900)$ & $(2100, 3000)$ & $(1060, 1400)$ & $( 670, 800) $\\
1000 &$(28700, 51200)$ & $(25700, 45200)$ & $(11100, 17900)$ & $(3450, 4950)$ & $(1760, 2400)$ & $(1160, 1450)$ \\
1500 &$(35200, 62600)$ & $(31600, 55600)$ & $(14100, 22700)$ & $(4500, 6500)$ & $(2400, 3200)$ & $(1650, 2100)$ \\
2000 &$(40700, 72300)$ & $(36600, 64400)$ & $(16600, 26800)$ & $(5520, 7950)$ & $(2990, 3950)$ & $(2100, 2560)$ \\
2500 &$(45400, 80700)$ & $(41000, 72100)$ & $(18900, 30500)$ & $(6450, 9250)$ & $(3550, 4700)$ & $(2550, 3100)$ \\
3000 &$(49800, 88500)$ & $(45000, 79150)$ & $(21050, 34000)$ & $(7300, 10500)$& $(4050, 5400)$ & $(2950, 3600)$ \\
\end{tabular}
\end{ruledtabular}
\end{table}

In conclusion, we put lower limits on $\Lambda_{\cal U}$ assuming
the scalar unparticle effects on the polarized cross section can
be distinguished from the SM contribution at $95\%$C.L. In our
calculations we consider the multi-TeV CLIC electron-positron
collider for the center of mass energies $\sqrt{s}=0.5$ TeV- $3.0$
TeV, and the luminosities ${\cal L}=100fb^{-1}$, and ${\cal
L}=1000fb^{-1}$ for different polarizations of the initial
electron, the laser photon, and the back-scattered  photon beams.
Our calculations show that the limits of $\Lambda_{\cal U}$
get stronger as one increases the
luminosity and the center of mass energy of the collider. It is
shown that, especially, for smaller values of the mass dimension
$d$, ($1 <d <1.3$), these bounds are very significant. As a
stringent limit, from TABLE \ref{tab1}, we find  $\Lambda_{\cal U}>~80$TeV for $d<1.3$ at
$\sqrt{s}=3$TeV and $1 ab^{-1}$ luminosity per year, which is
comparable with the limits calculated from other low and high
energy physics implications.

\section*{ACKNOWLEDGMENTS}
It is a pleasure to thank B. Balantekin for many helpful
conversations and discussions. KOO would like to thank to the
members of the Nuclear Theory Group of University of Wisconsin for
their hospitality, and acknowledges support through the Scientific
and Technical Research Council (TUBITAK) BIDEP-2219 grant. The
work of O. C. was supported in part by the State Planning
Organization (DPT) under grant no DPT-2006K-120470 and in part by
the Turkish Atomic Energy Authority (TAEA) under grant no
VII-B.04.DPT.1.05.

\end{document}